# Recommended Practices for Spreadsheet Testing

Dr. Raymond R. Panko
University of Hawaii
Panko@Hawaii.edu

## 1 ABSTRACT

This paper presents the author's recommended practices for spreadsheet testing. Documented spreadsheet error rates are unacceptable in corporations today. Although improvements are needed throughout the systems development life cycle, credible improvement programs must include comprehensive testing. Several forms of testing are possible, but logic inspection is recommended for module testing. Logic inspection appears to be feasible for spreadsheet developers to do, and logic inspection appears to be safe and effective.

## 2 INTRODUCTION

This paper presents recommended practices for spreadsheet testing. There is a major need for testing in order to reduce spreadsheet errors. At last year's EuSpRIG conference, the author summarized existing data on spreadsheet errors [Panko, 2005 2006b]. Among the findings are the following:

- ➢ Spreadsheets are widely used by many companies in financial reporting and other crucial business areas.

- ➢ Consistent with research on human error rates in other cognitive tasks, laboratory studies and field examinations of real-world spreadsheets have confirmed that developers make uncorrected errors in 2% to 5% of all formulas.

- ➢ Consequently, nearly all large spreadsheets are wrong and in fact have multiple errors. Field examinations of operational spreadsheets have confirmed this.

- ➢ In fact, field audits have found that a large percentage of spreadsheets have errors that are financially material [KPMG, 1988] or that could affect decisions [Coopers & Lybrand, 1997].

Consistent with human error research, which has seen similar error rates across cognitive tasks of comparable difficulty, spreadsheet error rates are very similar to those in traditional programming [Panko, 2005 2006b]. However, while programmers spend a quarter to 40% of their development time in testing Kimberland, 2004], testing among spreadsheet developers in industry is extremely rare [Panko, 2005 2006b].

Many things can be done to reduce the number of errors in spreadsheet development. However, the only proven way to reduce errors dramatically is testing. Programmers have learned that multiple rounds of testing can reduce error rates from 2% to 5% of all lines of code to only 0.1% to 0.3% [Putnam & Myers, 1992].

It would seem, then, that whatever else is done to reduce spreadsheet errors during development, comprehensive testing must be an ingredient in the mix. Consequently, testing is the focus of this paper. Unfortunately, although a great deal has been written about how to reduce errors in spreadsheet development, little has been written specifically about testing, and what has been written (e.g., Pricewaterhousecoopers, 2004) has little detail.

Note on terminology. The term "testing" means different things to spreadsheet developers and software developers. Spreadsheet developers use the term *testing* very broadly, to include looking over a spreadsheet informally, running an error checking program over the spreadsheet, auditing, execution testing, and logic inspection. Software developers have a much narrower view of *testing*. They reserve the name for what we will call execution testing—entering test data cases and determining if the program works as expected. They use other terms, such as inspection, for other post-development error reduction methods. We will use the term testing broadly, as it is used by spreadsheet developers.

This paper examines different types of testing that spreadsheet developers can use. It concludes that only logic inspection (an analog of code inspection) is likely to be suitable. It presents recommendations for how logic inspection should be done, based on Fagan's [1976, 1986] code inspection methodology or some other code inspection method.

## 3 TYPES OF TESTING

There are multiple types of testing, most of which are insufficient to reduce errors to the point where significant errors will be highly unlikely. The important criterion for selecting a testing methodology is the method's ability to reduce formula error rates from 2% to 5% to a far lower value, detecting 60% to 80% of all errors in each round of testing.

### 3.1 Testing During Development

Some SDLC phase models list development and testing as separate phases, with testing beginning after development ends. Actually, testing takes place throughout the development process.

**Requirements Testing**

In fact, testing should begin well before code or logic creation. As discussed later, code/logic inspection can be done on requirements documents and other early-stage documents. Many errors are introduced before coding ever begins, and testing specification documents and other documents can reduce these errors.

**Unit Testing**

After a developer has finished a module and examined his or her work for errors, the module must be subjected to unit testing. Modules are small enough to make them easily testable. As discussed below, modules are small enough for code inspection to be used to test them.

**Integration Testing**

After modules are tested, the modules are integrated into larger units. Usually, several stages of integration are needed, each with its own stage of testing. Although integration testing is very important, we will focus on testing techniques for unit testing.

### Agile Development Methods

Our discussion of phases has assumed a traditional software development life cycle (SDLC) model. However, just as there are many development methodologies for software, especially new agile methods, spreadsheet development may be done in non-traditional ways, especially agile methods. Different development disciplines require different testing phases.

### 3.2 Eyeballing

One "testing" technique is looking over the spreadsheet for reasonableness. This often is called eyeballing or squinting. Nardi and Miller [1991] found that having a colleague check a spreadsheet for reasonableness is very common. In fact, it seems to be a good way to *begin* testing a spreadsheet. However, there is no evidence that eyeballing substantially decreases error rates. Nor is there any evidence for the frequent statement that eyeballing will catch *the most serious* errors. In fact, the limited research that has been done in error recognition suggests that people are poor at recognizing even serious errors [Klein, Goodhue, & Davis, 1997; Rickets, 1990].

### 3.3 Error Scanning Software

Another tool for initial error discovery is error scanning software, which searches through the spreadsheet seeking errors based on discovery rules.

### Analogy: Spell Checking and Grammar Checking

Automated error scanning tools are similar to spell checking and grammar checking tools in word processing programs. Although spell checking and grammar checking tools are important in word processing, experienced writers know that they only catch a fraction of all errors.

In addition, just because a grammar checker detects an error correctly, this does not mean that the user will accept the suggestion. In general, errors scanners merely identify potential problems. Only if the user accepts a suggestion does correct detection matter.

Also, grammar checkers and spell checkers often produce false alarms, in which a flagged piece of text actually is correct. When Galletta, et al. [2005] had students write with or without a grammar checker, they found, to their surprise, that students who used grammar checkers actually made *more* grammatical errors than students who did not. Closer analysis of the data showed that students who used grammar checkers frequently accepted incorrect suggestions.

When we look at automated error detection programs, the same issues are likely to appear: errors missed by the checker programs, errors correctly identified by the checker but dismissed by the user, and user acceptances of incorrect suggestions.

### Error Checking Tools

Excel 2003 has a built-in error checking tool. Under the Tools menu, select Error Checking. If the Excel Error Checking tool detects a possible error, it pops up a dialog box. This box gives several choices, which range from ignoring the warning to taking action. The error checking tool built into Excel is simple but limited. A number of more sophisticated error checking products are on the market. A good list of these products exists at Patrick O'Bierne's Sysmod.com website. The page showing the list of products is http://www.sysmod.com/sslinks.htm#auditing.

**Are Spreadsheet Error Checkers Safe and Effective?**

Spreadsheet error checking products are attractive because they can quickly check a spreadsheet for errors. However, to adopt a medical expression, are these products safe and effective?

Effectiveness means that they work *well*. If error checking programs are to be used only as initial error scanners, there is no problem. However, if they are to be used as main testing tools, they must catch nearly all errors or at least a large fraction of all errors.

For example, error checking tools are not effective as full testing tools if they are blind to major categories of errors. For instance, in omission errors, something is left out of the model. If there are ten important cost items for a company, and if the developer forgets to put in a row for one of these cost items, this is an omission. It is difficult to see how an error checking program could detect this. (Proponents of error checking programs have suggested that there will be a blank row, but this is unlikely if the item was simply forgotten during development, unless you believe in subconscious blank row insertion.)

Another problem exists if the developer has the wrong algorithm for a formula or if the developer expresses the formula incorrectly. In cases like this, there is not likely to be anything in the formula's expression to indicate that it is incorrect.

The other issue is safety. As noted earlier, the Galletta et al. [2005] study showed the danger of false alarms being accepted by users. Again, empirical testing is necessary to determine how serious the problem is in practice.

Overall, the fraction of all errors that an error checker finds is a question for empirical analysis. Again, however, that this is only an issue if the error checker is expected to be a full testing tool. The author would not argue that automated error checking should be avoided—only that expectations of its effectiveness should be modest.

## 3.4 Auditing

One source of great confusion in the spreadsheet literature is the term "auditing." Auditing is a term that came out of internal auditing, which normally focuses most heavily on financial auditing.

In auditing, the auditor does not examine everything. Rather, the auditor asks many questions whose answers may indicate problems. In addition, the auditor does spot checking instead of examining every item (transaction, asset, etc.). The goal of auditing is *emphatically not to reduce errors*—merely to detect indications of problems.

Adding to the confusion, some published "audit" studies of spreadsheets are really full logic inspection studies. This includes the widely cited "audits" by Galletta, et al. [1993, 1997]. We will refer to these studies again later, under logic inspection.

One problem with true auditing is the issue of what parts of the spreadsheet to audit. An obvious candidate is, "the riskiest parts," including complex formulas and links between worksheets in a workbook. However, while these types of formulas are indeed more likely to contain errors than average formulas, there probably are tens of average formulas for every obviously error-prone formula. Error has a strong random component, so there are likely to be many more errors in total in average formulas than in complex formulas.

Overall, spreadsheet audits (in the noncomprehensive traditional sense) are best viewed as another category pretesting tools, like automated spreadsheet error checkers. However, while error checking programs are simple, fast, and inexpensive to use, spreadsheet audits are complex and expensive. It is not clear why audits should be undertaken if the firm has a serious commitment to reduce errors to acceptable levels though comprehensive testing.

**3.5 Execution Testing**

To programmers, "testing" means what we will call execution testing. In execution testing, the tester tries several suites of input values to see if they produce (or do not produce) correct results.

Execution testing is very widespread in software development, and it may seem that it would be straightforward to do. In fact, however, execution testing requires extensive training to do well.

**Selecting Input Value Suites that May Break the Code**

One issue is how to select input value suites. Common advice in recommended spreadsheet practices is to "use both typical and extreme values." This is simple. It is also wrong—or at least insufficient.

One problem is that most people have a strong a confirmation bias, which causes them to pick values that are likely to prove that their code is correct. Confirmation biases have been seen in many contexts besides testing [Wason, 1960].

Testers must fight this natural tendency and develop the habit of picking input value suites that may or should break the code. For counting variables, they should try negative values, text values, and other nonsensical values. Although these seem ludicrous, in practice users often input odd values through confusion. In addition, "impossible" values often reveal problems that are likely to cause troubles sporadically or that make a program susceptible to denial-of-service or hacking attacks.

**Paranoid Testing**

Selecting input values normally is done assuming innocent errors. When security is an issue, however, the tester must conduct paranoid testing in which he or she inputs values that reflect not only innocent errors but also malicious input values that could be used by attackers or fraudulent employees. For instance, many database programs are susceptible to SQL injection attacks in which the user enters an SQL statement instead of an expected value. For instance, when prompted for a password, the user might enter "Dumbpassword OR TRUE". If the program parses this SQL query, the result is automatically TRUE, and the attacker will be authenticated improperly.

**Groups of Input Value (Suites)**

Even if people can be trained to select values for single input values intelligently, execution tests almost always involve input value *suites* that specify test values for several input variables. Execution testers must be trained in how to select combinations of input values effectively because failures often occur because of subtle combinations of input values. Selecting values for suites of input variables is even more difficult than selecting values for single variables.

If there are more than two or three input variables in the suite, the number of possible input value combinations becomes enormous. This is called the combinatorial explosion. To reduce this problem, testers must learn to create equivalence classes—single value suites that can stand for a wide range of value suites. For instance, to test a table lookup in a spreadsheet formulas, testing a relatively few values may be sufficient, although the tester must be wise enough to test out-of-range lookups and end-of-range lookups.

**Code Coverage**

In programming, code coverage is a common input suite selection concern. Due to the presence of many branching statements, it is difficult to select sets of input value suites that will test all lines of code. In fact, it is nearly impossible to do so. Testers learn to create reasonable levels of code coverage instead of 100% code coverage, but not testing all possible paths is always risky. In spreadsheets, there often are only a limited number of paths through the spreadsheet's formulas. However, the use of IF functions is fairly common and creates branches in the spreadsheet execution—sometimes many branches.

**The Oracle Problem: The Big Issue**

In execution testing, various inputs are tried to see if they produce correct results. However, how does one know when a result is correct? This is called the oracle problem. An oracle is way of knowing when results are correct (more precisely, when the results are the expected ones). For example, when testing a transaction processing program that creates new accounts, the tester can enter data for a new individual and then check to be sure that a new account was created and that it contains all of the input values. Many oracles, like this one, are obvious.

For spreadsheets, there seldom are obvious oracles such as crashes or other obvious failures. Instead, the results of calculations are incorrect. How can one know whether the results of tests are correct? A common bromide is to say that testers should compare the results with known calculated values. However, in most cases, there was no comparable calculation before spreadsheets. In nearly all spreadsheets, calculations are extended well beyond what had been done previously with manually calculations. In complex spreadsheets, then, there usually is no oracle other than the spreadsheet calculations, which may not be correct.

It could be argued that one way to create a good oracle is to create a mathematical model before building the spreadsheet. The model could then act as an oracle. However, people also make errors building mathematical models, and the only way to calculate a mathematical model may be to express it in a spreadsheet. Although creating mathematical models does seem promising, its appropriateness and limitations need to be studied empirically before being accepted.

This lack of a readily-found oracle probably is the most serious problem in spreadsheet execution testing. Without a strong and easy-to-apply oracle, execution testing simply makes no sense for error-reduction testing even if training problems could be overcome.

**Regression Testing**

However, even in spreadsheet development execution testing may be attractive for regression testing, which is undertaken after a modification to ensure that it still works properly in parts where no changes should result. Value suites previously used in the model are entered into the model, and the "bottom-line" figures are compared to their values before the spreadsheet was changed. In regression testing, the fact that the only realistic oracle is the spreadsheet itself is not a problem.

**Perspective**

Execution testing seems attractive. Plug in input values and see if the output is OK. However, selecting test values is an exacting discipline, and it takes professional testers weeks to training to be effective. It seems unrealistic to expect spreadsheet developers, who nearly always have many professional responsibilities that have nothing to do with spreadsheets or even IT, to take the time out to receive training. Worse yet, the result of spreadsheet calculations is numerical results, and spreadsheets usually are so complex that no realistic oracle exists. For spreadsheet testing, execution testing does not appear to be promising, and companies that decide to use it need a plan to address the limitations of execution testing.

### 3.6 Logic Inspection

The last testing method we will cover is also the one that we believe to be a good recommended practice. This is logic inspection, which is called code inspection in software development.

In code inspection, inspectors examine a program module, line by line, looking for errors. In logic inspection for spreadsheets, inspectors examine all formula cells. Note the words "entire" and "all" in the previous two sentences. In contrast to auditing, which frequently focuses only on parts that seem risky, logic inspection is comprehensive in what it examines.

Code inspection is attractive because it requires far less training than execution testing (although in the next section we will see that considerable discipline must be used in code inspection).

Finally, as well as being a good way to do testing during the coding and testing phase, logic inspection can also be applied during the earlier requirements and design phases in development. This allows errors to be detected before coding even begins, when the cost of fixing an error is relatively small, and when most errors are generated. Execution testing cannot used for this important early phase testing.

### 4 LOGIC INSPECTION PRINCIPLES

Having identified logic inspection as the most promising testing approach for spreadsheet developers to use to greatly reduce errors, we will look at how logic inspection should be done. Although it may seem that logic inspection is simple ("Just look at everything"), logic inspection is a highly refined processes.

### 4.1 The Need for Team Inspection

In software development, Fagan [1976, 1986] developed the code inspection methodology at IBM. Although there are several variants, we will discuss Fagan inspection as a basic model for spreadsheet logic inspection.

Fagan inspection mandates *team inspection* rather than individual inspection. The reason for team inspection is pragmatic; individuals typically catch fewer than half of all errors in real-world code inspections. This finding has been repeated (and explored in more detail) in laboratory experiments. In these experiments in software code inspection, subjects usually only catch 40% or fewer of all defects [Basili & Selby, 1986; Johnson & Tjahjono, 1997; Myers, 1978; Porter, Votta, & Basili, 1995; Porter & Johnson, 1997;

Porter, Sly, Toman, & Votta, 1997; Porter & Votta, 1994]. Having multiple inspectors raises this detection rate substantially. For instance, in the Johnson and Tjahjono study [1997], individuals caught only 23% of all seeded errors, while the average team of three caught 44%. Even the best group only found 63% of the seeded errors.

The need for team logic inspection has also been demonstrated in the laboratory for spreadsheet testing. Panko [1999] had students work alone to detect errors and then meet in teams of three to pool their results and find new errors. On average, individuals caught 63% of all errors, while teams of three caught 83%. Although groups of three only caught about a third more errors than individuals, the gains from team development came precisely from the errors that individuals had the most difficult time finding.

More generally, Steiner [1972] summarized small group research, including groupwork and its relationship to error reduction. His summary found that the gains from team inspection seen in programming and in Panko's [1999] spreadsheet logic inspection study are unsurprising in view of teamwork's long-proven ability to reduce errors more than individual work.

**4.2 Having Reasonable Expectations**

Note that code inspection does not entirely eliminate errors. Error-reduction methods that use humans never eliminates all errors in software development, grammar or spelling [Panko 2006a], or any other aspect of life. Typically, as just noted, a round of team code inspection is likely to eliminate 60% to 80% of all errors in software.

We have noted Panko's [1999] laboratory results for group spreadsheet logic inspection. Earlier, two studies by Galletta, et al. [1993, 1997] explored error detection by individuals but not by groups. In the 1993 study, subjects detected 56% of errors on average. CPAs (not CPA students) caught slightly more errors than did MBA students, but experienced spreadsheet users did not find more errors than inexperienced spreadsheet users.

**4.3 Team Roles**

In Fagan inspection, there are two distinct roles on the inspection team.
- ➢ The central role is that of the moderator, who manages the process.
- ➢ Then, of course, there are testers, who do the actual inspection.

In addition, some testers are likely to take on specialized roles during the inspection meeting.
- ➢ One of these testers will be the developer of the spreadsheet. Although it is often said that people are not good at catching their own errors, research in proofreading does not support this belief [Daneman & Stainton, 1993]. However, Reason [1990] has shown data that individuals have a difficult time recognizing omission or misdiagnosis errors.
- ➢ One of the testers is likely to be designated as the reader, who paraphrases the spreadsheet logic being discussed.
- ➢ One of the testers may be the recorder, who records spreadsheet errors found during the inspection meeting.

### 4.4 Process Phases

Fagan's inspection process has seven steps.

- ➢ Planning. Preparing materials, obtaining participants, and scheduling meetings.
- ➢ Overview Meeting. Introduces the software, lays out roles, and describes process.
- ➢ Preparation. Inspectors, working alone, examine the spreadsheet. According to Fagan, the goal is NOT to identify errors but to understand the software module. Many inspection methods, however, do engage in error hunting during preparation.
- ➢ Inspection Meeting. The goal is to detect errors. There should not be discussions beyond those needed to identify and clarify errors. Meetings must be kept reasonably short to prevent loss of vigilance.
- ➢ Process Improvement. The inspection also should provide feedback to the firm's inspection process guidelines. Each inspection must generate statistics on time spent, errors found, and error severity.
- ➢ Rework. Fixing the software is done after the meeting.
- ➢ Follow-Up. To ensure that changes are made appropriately.

### 4.5 Parameters

In the previous discussions, we have noted that teams must be used and that there must be time limitations to prevent the loss of diligence. In this section, we will look at specific suggestions for key parameters.

**Module Size**

A general rule in code inspections is that preparation and inspection should be limited to two hours. Modules typically consists of 200 to 400 lines of code [Panko, 2006a], so preparation and inspection rates must be approximately 100 to 200 lines per hour (LPH). Surprisingly, module size has not been varied systematically in experiments, although Barnard and Price [1994] reported that inspectors in the code inspections they knew found 72% more errors when modules were smaller.

**Basic Speed**

Greater attention has been given to preparation and inspection rates.

- ➢ Basili & Perricone [1993] found that when preparation and inspections were done at 50 LPH, inspectors found errors in 1.6% of all lines of code. This fell by 25% to 1.2% when the speed was tripled to 150 LPH and fell another 50% to 0.6% when the inspection rate increased to 200 LPH. Between 50 LPH and 200 LPH, the error detection rate fell by 62%.
- ➢ Russell [1991], in turn, measured a 3.7% error detection rate at 150 LPH, 1.5% at 450 LPH, and 0.8% at 750 LPH. The increase from 150 LPH to 750 LPH brought a 78% reduction in error detection.
- ➢ Ebenau &Strauss [1994] did not report specific preparation or inspection speeds, but they noted that "non-hasty" inspection yielded error detection rates of 2.0%, while "hasty" inspection yielded error detection rates of 1.3%—a 35% reduction.

**Team Size**

For code inspection, typical team sizes are three or four. For document inspection, teams usually are slightly larger. There has been surprisingly little research on team size. Only one study has looked specifically at team size. Weller [1993] looked at error detection rates for teams of three and four. For teams of three, the error detection rate was 1.8%. This rose to 2.8% for teams of four—a gain of a little over half.

Weller's [1993] study compared both team size and inspection rates. As Figure 1 illustrates, Weller looked at groups of three and four testers and compared preparation rates of less than 200 LPH and more than equal to 200 LPH. Note that error detection rates varied considerably despite the relatively small differences in team sizes. For inspection speeds greater than 200 LPH, increasing group size from three to four *doubled* the error yield.

*Figure 1: Error Detection Rates by Team Size and Preparation Rate*

|            | <200 LPH | >= 200 LPH | Both |
|------------|----------|------------|------|
| Team of 3  | 2.4%     | 1.2%       | 1.8% |
| Team of 4  | 3.1%     | 2.5%       | 2.8% |
| Both       | 2.8%     | 1.9%       | 2.3% |

Source: Weller [1993].

## 5 RECOMMENDED TESTING PRACTICES FOR SPREADSHEETS

### 5.1 Recommendation 1: Create a Policy Requiring Comprehensive Testing

The first recommendation is that companies must have policies requiring comprehensive spreadsheet testing for all important spreadsheets. This comprehensive testing must consume 25% to 40% of all spreadsheet development time.

> Without comprehensive inspection, there will be errors in roughly 2% to 5% of all root (non-copied) formulas, nearly all large programs will contain many errors, and most programs will have significant errors (material financial errors, errors large enough to influence decisions, and so forth). In addition, human beings are not very good at detecting errors. It is important to have correct error beliefs because if errors are underestimated, companies cannot develop appropriate policies.

### 5.2 Recommendation 2: If pretesting is done, looking over spreadsheets for errors and error checking programs are good but should not be used in lieu of comprehensive testing.

Second, looking over spreadsheets for errors and using error checking programs are good practices for pretesting but do not constitute effective comprehensive testing. Auditing ("only looking over the risky parts") is only a pretesting method but is controversial because it is quite expensive and checks parts that will have to be checked later anyway, during comprehensive testing.

### 5.3 Recommendation 3: Only use execution testing for regression testing unless the user training and oracle problem can be overcome.

Third, execution testing constitutes comprehensive testing, but it should not be used unless companies can be sure that they can address two major problems with spreadsheet execution testing—the need to train users in how to develop input value suites and the difficulty of finding an oracle. In regression testing, the oracle problem is solved, making execution testing more attractive. If execution testing is done, it needs to be done on modules to reduce the combinatorial explosion problem.

### 5.4 Recommendation 4: Use logic inspection for comprehensive testing.

Fourth, logic inspection is recommended for comprehensive testing because ordinary spreadsheet developers probably can employ it effectively, because a round of logic inspection can detect 60% to 80% of all errors, and because there are proven methodologies for doing logic inspection. We have discussed the Fagan method, but there are quite a few other forms of "formal technical review." It is recommended that inspection be done on spreadsheet modules and also on documents created before coding (requirements documentation, etc.) and after coding (logic documentation, etc.)